\documentclass[11pt, a4paper]{article}
\usepackage[utf8]{inputenc}
\usepackage[UKenglish]{babel}
\usepackage[textwidth=16cm]{geometry}
\usepackage[center]{caption}
\usepackage{graphicx}
\usepackage{hyperref}
\usepackage{amsmath}
\usepackage{natbib}
\graphicspath{{./pictures&movies/}}
\usepackage[symbol]{footmisc}
\usepackage{lineno}

\title{Saving Energy in Turbulent Flows with Unsteady Pumping}
\author{Giulio Foggi Rota$^{1,2}$, Alessandro Monti$^1$, Marco E. Rosti$^{1*}$, Maurizio Quadrio$^{2*}$}
\date{$^1$ Complex Fluids and Flows unit, Okinawa Institute of Science and Technology Graduate University, 1919-1 Tancha, Onna-son, Okinawa 904-0495, Japan \\
$^2$ Dipartimento di Scienze e Tecnologie Aerospaziali, Politecnico di Milano,
via La Masa 34, 20156 Milano, Italy\\
$^*$Corresponding authors: marco.rosti@oist.jp, maurizio.quadrio@polimi.it}
\begin{document}
\maketitle

\section{Summary}
Viscous dissipation causes significant energy losses in fluid flows; in ducts,  laminar flows provide the minimum resistance to the motion, whereas turbulence substantially increases the friction at the wall and the consequent energy requirements for pumping.
Great effort is currently being devoted to find new strategies to reduce the energy losses induced by turbulence. 
Here we propose a simple and novel drag-reduction technique which achieves substantial energy savings in internal flows.
Our approach consists in driving the flow with a temporally intermittent pumping, unlike the common practice of a constant pumping.
We alternate ``pump on" phases where the flow accelerates, and ``pump off" phases where the flow decays freely. 
The flow cyclically enters a quasi-laminar state during the acceleration, and transitions to a more classic turbulent state during the deceleration.
Our numerical results demonstrate that important energy savings can be achieved by simply modulating the power injection into the system over time.
The physical understanding of this process can help the industry in reducing the waste of energy, creating economical benefits and preserving the environment by reducing harmful emissions.

\section{Introduction}
The world is facing an unprecedented energy problem \cite{gray-2017}.
Meeting the increasing demand for energy while reducing our dependence from fossil fuels is a complex and multi-faceted problem, challenging researchers \cite{murshed-tanha-2021,luderer-etal-2022} and international organisations like the International Energy Agency \cite{IEA-2021}.
While some countries are able to trade part of their wealth for a more sustainable energy production, others do not have such a possibility.
Due to the recent growth of the oil price, in facts, the latter are left with the only option of dramatically increasing their reliance on firewood. 
Such trend clearly contrasts with the guidelines drawn by the Glasgow Climate Pact \cite{glasgow-pact}, enhancing both deforestation and global warming.
In order to interrupt this escalation, it is necessary to reduce the price of liquid fuels. 
Cheaper access to natural oil and gas can facilitate the economic growth of developing countries, later allowing them to face the transition towards greener energy sources.
At the same time, a lower cost of innovative fuels, like hydrogen, can further promote the transition, also making it easier for those nations already undergoing it.  

Various factors contribute to the final cost of fuels, as frequently discussed in heterogeneous contexts, but the non-negligible share coming from transport is often overlooked \cite{masnadi-etal-2021}.
As recently estimated \cite{saady-lewis-mcfarland-2018}, the cost of energy transport may vary by more than two orders of magnitude when comparing different carriers and energy sources.
Transferring oil and gas in tankers, for example, is significantly more convenient than displacing liquid hydrogen in pipelines. 
Oil and gas, in fact, have a significantly higher energy density than liquid hydrogen, therefore less fluid needs to be transported in order to store the same amount of energy.
Nevertheless, all these fuels remain unaffordable for many developing countries, and their transport cost needs to be reduced.

The sizes and operative flow-rates of conventional pipelines determine an highly turbulent motion of the fluid at their interior. 
Its chaotic behaviour dissipates a significant amount of the energy supplied to the flow intended to displace it, hence inducing significant drag penalties and decreasing the energetic efficiency of the process.
Over the years, many researchers have studied  various approaches to control the flow \cite{bushnell-moore-1991,lumley-blossey-1988,gadelhak-2000,kuhnen-etal-2018} and have categorised them as active or passive, depending on whether they require a power supply.
Special surface treatments such as the carving of microscopic riblets on the wall \cite{garcia-jimenez-2011-a} or the addition of polymers to the flow \cite{virk-1975} belong to the second category, while for example a predetermined motion at the wall \cite{du-karniadakis-2000,quadrio-2011} belongs to the first. 
Some techniques are extremely efficient in the context of duct flows (e.g. polymers addition can reduce the drag up to 80\%, and some moving-wall strategies approach flow relaminarization), but most of them bring disadvantages like the contamination of the fluid with spurious substances, the introduction of complex control mechanisms with moving parts, or the need for regular maintenance. 

In this paper, we demonstrate the success of a simple and feasible approach that circumvents such issues. The periodic power-on and shutdown of the pump feeding a turbulent pipe flow can reduce the energy needed to transfer a fixed amount of fluid in a given time from one point to another, regardless of the power peaks required to repeatedly accelerate the fluid. 
This constitutes a hybrid form of control, in which no additional devices are required (as in passive techniques), and the active pumping phase is employed to increase the overall efficiency of the system (as in active techniques).
Building upon recent progresses made in understanding the transient nature of turbulent flows and the effects of a time-varying pumping \cite{mathur-etal-2018, jovanovic-2021}, our approach exploits an unsteady power delivery, inspired by the work of Iwamoto, Sasou and Kawamura \cite{iwamoto-sasou-kawamura-2007} and Kobayashi et al. \cite{kobayashi-etal-2021}, designed to move the flow smartly back and forth between the turbulent regime and a quasi-laminar one.
In the following, we describe in detail our innovative forcing technique, and discuss the results of the demanding numerical simulations we performed.
The potential of this control and the need for further investigations are highlighted.

\section{The unsteady pumping}

Our numerical experiments are carried out in an indefinite channel, bounded by two smooth planar walls.
An incompressible Newtonian fluid flows through the domain as described by the incompressible Navier--Stokes equations, under the influence of a volumetric forcing of choice. 
In such simulations, this forcing is either set to a constant pressure gradient, or adjusted in time to ensure a constant flow rate. 
Here, instead, we prescribe an homogeneous streamwise pressure gradient $\Pi(t)$ which periodically switches between a constant value and zero, as shown in panel $a$ of figure \ref{fig:0D}, where all quantities are compared to those of the reference flow, indicated with the subscript 0. This is a conventional turbulent channel at $Re_{m,0} = U_{b,0} 2h / \nu = 5626$ (or $Re_{\tau 0} = u_{\tau 0} h / \nu = 180$) as that of Kim, Moin and Moser \cite{kim-moin-moser-1987}, where $U_b$ represents the flow rate per unit cross-sectional area of the channel, $h$ its half-width, $u_\tau$ is the friction velocity, and $\nu$ the kinematic viscosity of the fluid.
Restricting the average value of the on-off pressure gradient to be equal to the reference one makes its waveform depend on two parameters: the pulsation period $T$ and the relative length of the active pumping phase, the duty-cycle $\xi$.
Formally, for the $n^{th}$ pulsation we prescribe: 
\[
\frac{\Pi(t)}{\Pi_0} = 
\begin{cases}
\begin{aligned}
 \frac{1}{\xi} \qquad & \mbox{for} \quad n \le \frac{t}{T} < (n+\xi) , \\
 0             \qquad & \mbox{for} \quad (n+\xi) \le \frac{t}{T} < (n+1)
\end{aligned}
\end{cases}
\]

A similar control strategy was recently studied experimentally by Kobayashi et al. \cite{kobayashi-etal-2021}, who employed machine learning to devise and test different waveforms. In our numerical study, however, we employ a simpler and more realistic waveform, and aim at verifying the robustness of the approach with respect to the investigation method and the control parameters, establishing that the advantage is not limited to reductions in drag, but extends to savings in energy.
  
Within this study, we have considered different values of the period ($T \in$ \{310, 930, 1245, 1555\}, expressed in units of the convective time scale $h / U_{b,0}$) and of the duty-cycle ($\xi \in $\{0.1, 0.05, 0.0375, 0.025, 0.0125, 0.005\}).
To appreciate the computational challenge implied by the present work, one should consider that $T=1555$ is approximately fifty times longer than the time computed by Kim, Moin and Moser  \cite{kim-moin-moser-1987} to obtain reliable statistics of the channel flow at the same Reynolds number. Moreover, for each $T,\xi$ combination, several tenths of pumping cycles need to be simulated to properly asses the energy savings (see Supplementary Material), and millions of CPU hours are needed. 
This is, nevertheless, the only way to capture numerically the true physics of the flow (much like in a laboratory experiment), resolving all the scales of motion without resorting to any kind of model.

An extensive investigation of the parameter space to find the optimal performance is beyond the aims of this paper: instead, we seek to establish the fundamental result that a temporal modulation of the pumping power can produce a net energy benefit. 
To this purpose, even though the control proved effective for all the $(T,\xi)$ couples associated to the highest values of period and duty-cycle, in the following we will consider only the best performing case identified so far.

\section{Results}
The best performance to date has been found for the longest values of period and duty cycle, i.e. $T=1555$ and $\xi=0.1$. The energy spent to transfer a unit mass of fluid through the channel undergoes an outstanding reduction of 22\% of the energy requirement in an uncontrolled scenario with the same flow rate.
From a different perspective, the time employed by a unit mass of fluid to travel for a prescribed length is decreased by 13\% of the time required in an uncontrolled scenario with the same energy expenditure.
These figures, being integrated over the pumping cycles, account for the cost of periodically accelerating the fluid. Further details are reported in the Supplementary Material.

\begin{figure}[h!]
\centering
\includegraphics[scale=0.35]{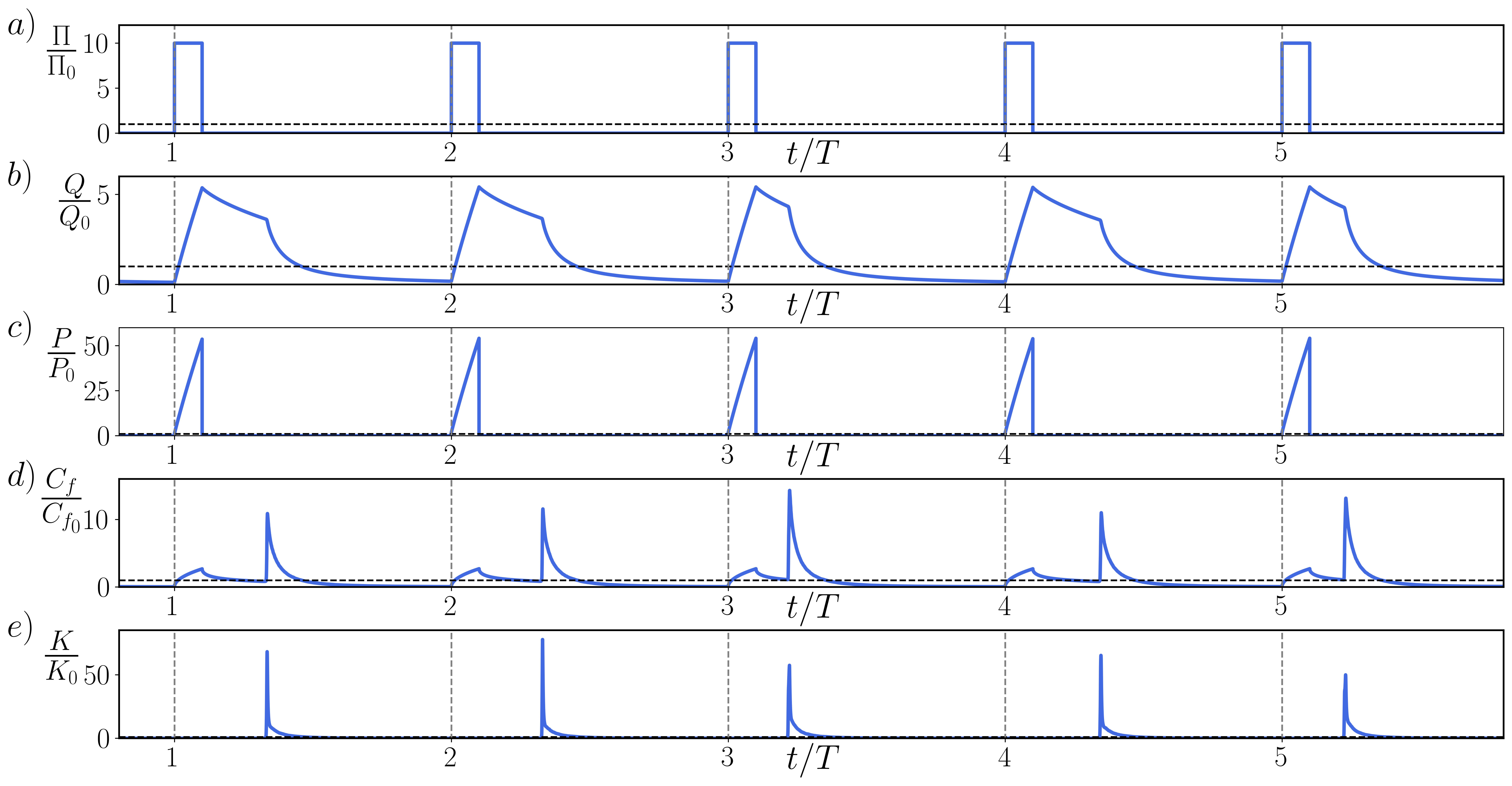}
\caption{\textbf{Time history of the flow performance:}
we report the evolution of ($a$) the pressure gradient $\Pi$, ($b$) the flow-rate $Q$, ($c$) the pumping power $P$, ($d$) the skin-friction coefficient $C_f$ and ($e$) the cross-stream turbulent kinetic energy $K$ over five periods. Those are chosen from a simulation with parameters $T=1555$ and $\xi=0.1$, which yield the maximum energy savings of our study. All quantities are referred to those of an uncontrolled turbulent channel flow at  $Re_{m_0} = 5626$.
During the active pumping phase the flow-rate and the pumping power increase.
After the pressure gradient is switched off (and the pumping power consequently becomes zero), the flow temporarily lingers in a quasi-laminar state, with high flow rate and low skin friction.}
\label{fig:0D}
\end{figure}

Figure \ref{fig:0D} shows the time history of global flow quantities.
Different periods are separated by vertical dashed lines, while horizontal lines mark the unitary value of the reference.
As soon as the driving pressure gradient $\Pi(t)$ is switched on, the flow undergoes a severe acceleration.
The flow rate (denoted $Q$ in panel $b$) starts to grow and the pumping power (denoted $P=U_b \cdot \Pi$ in panel $c$) increases accordingly.
The wall friction (denoted $C_f$ in panel $d$) surges as well.  
Greenblatt and Moss \cite{greenblatt-moss-1999} noticed for the first time how, during an intense acceleration, turbulence may be destroyed and the flow moves towards a quasi-laminar state.
We therefore observe the evolution of the turbulent kinetic energy (denoted $K$ in panel $e$) associated to the cross-stream velocity fluctuations: an efficient indicator of the turbulent state of the flow.   
The value of $K$ is extremely small at the beginning of every period, and remains such during the acceleration. 
The forcing is then switched off setting $\Pi=0$; from now on, no power is needed to drive the flow.
$Q$ starts decreasing (due to viscous losses) and so does $C_f$, while there is no significant variation in $K$.
At a random instant of the deceleration, however, the decay rate of $Q$ suddenly increases, producing a kink in the curve of panel $b$.
This coincides with a peak in $K$ and, shortly thereafter, also $C_f$ reaches its maximum: the flow moves from a quasi-laminar to a fully turbulent regime.
The later this transition occurs, the larger the cycle-averaged value of the flow rate and thus the effectiveness of the control.
After becoming turbulent, the flow undergoes a regular decay until the end of the period, where $Q$ and $C_f$ are low and $K$ is again almost zero.

\begin{figure}
\centering
\includegraphics[scale=0.35]{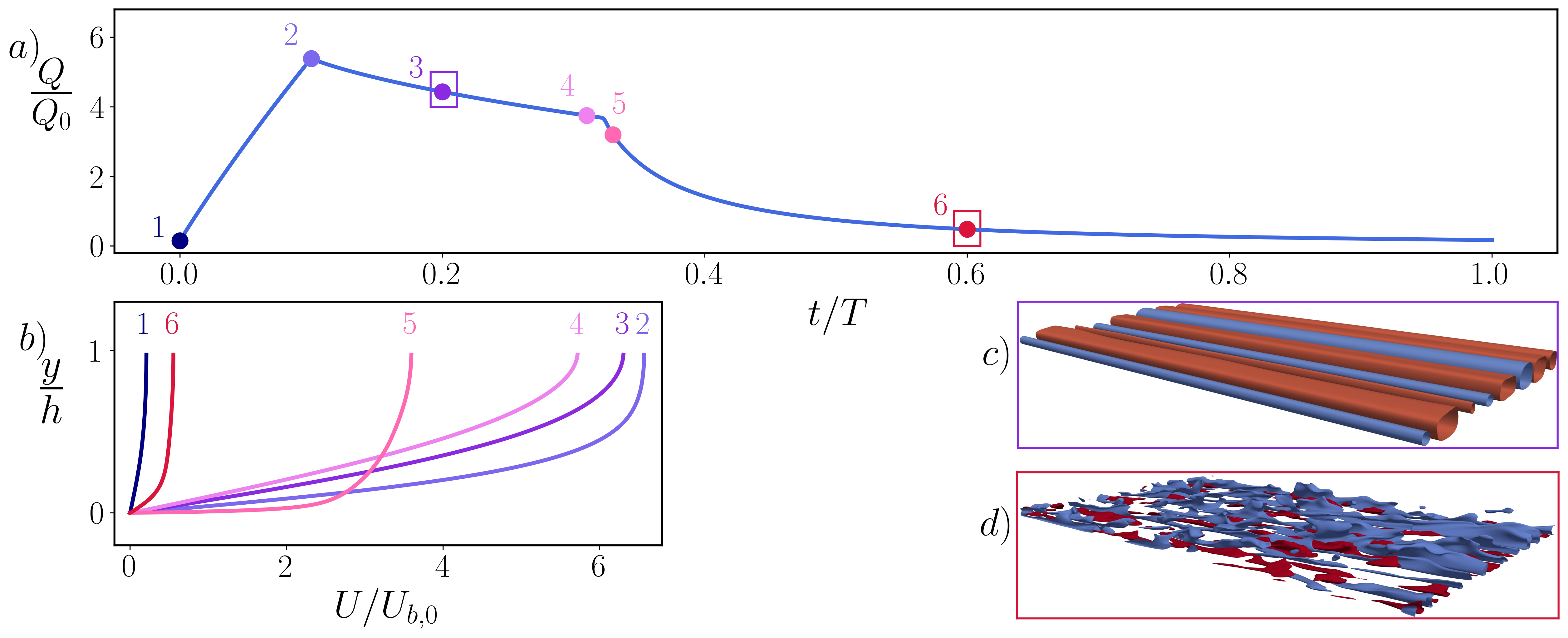}
\caption{\textbf{Flow dynamics over one cycle:}
we report ($a$) the evolution of the flow rate $Q$ over one cycle and ($b$) the mean streamwise velocity profiles $U$ at six selected time instants of the cycle, identified by  numbered dots in panel $a$.
For two time instants ($3$ and $6$) we also plot, in panels $c$ and $d$ respectively, the negative (blue) / positive (red) iso-surfaces of the streamwise velocity fluctuations ($u$) in half of the channel, i.e., $u/U_{b,0}=-0.1, 0.04$ ($c$) and $u/U_{b,0}=\pm 0.1$ ($d$). The quasi-laminar flow phase is ruled by intense and anomalous velocity streaks ($c$), which eventually become unstable and cause the breakdown to turbulence ($d$).}
\label{fig:period}
\end{figure}

In figure \ref{fig:period}, we illustrate the global flow dynamics over one cycle.
Panel $a$ represents once again the flow rate, while dots refer to the instants where space-averaged streamwise velocity profiles are plotted in panel $b$.
Wall friction builds up while the flow rate grows, as indicated by the increase in the wall slope of the velocity.
Meanwhile, the classic low- and high-speed streaks \cite{kline-etal-1967,smits-mckeon-marusic-2011}, which extend towards the centre of the channel before the activation of the forcing, are stretched and initially stabilised. 
The sudden, intense and brief nature of the acceleration prevents the appearance of turbulent spots, and no new turbulence is therefore generated during this phase.
At the end of the acceleration, what remains of the classic streaks is a quasi–2D perturbation (in the spanwise / wall-normal plane) of the laminar flow: the streaks extend indefinitely in the streamwise direction. 
After switching the forcing off, they do not become immediately unstable, but rather keep growing while the flow progressively slows down, and organise themselves in a regular alternating low- and high-speed pattern (as in panel $c$), which sometimes extends up to the centreline, while the mean velocity profile assumes the typical parabolic shape reminiscent of the laminar regime.
Such streaks differ from those of a conventional turbulent channel flow with comparable wall friction, and are observed here for the first time.
This picture comes to an end when a flow instability sets in and destroys the streaks, making the flow turbulent as in panel $d$: perturbations suddenly grow in magnitude, and an abrupt transition drives the channel towards a fully chaotic regime.

Once new turbulence is generated, the wall friction undergoes a sudden increase: this induces a faster decay of the flow rate, corresponding to the kink in figure \ref{fig:0D}, while the mean velocity profile assumes a plug-like shape.
It must be stressed how the temporal position of the kink varies randomly among different periods: the energy saving is determined by its cycle-averaged value, and this motivates the need to simulate several cycles to obtain a reliable estimate of the saving.
After transition, the turbulent flow undergoes a classic decay until the beginning of the next acceleration phase (videos of the transition and the turbulent decay shown in figure \ref{fig:period} are provided in the Supplementary Material).

\section{Discussion}

Our work demonstrates how the energetic efficiency of a fluid transport system can be effectively improved by modulating in time the pumping power. 
By employing a simple periodic sequence of on--off pulses, described by the duration of the period and the duty-cycle, our numerical study proves that intense accelerations followed by long relaxation phases yield significant energetic savings, up to 22\% in the parameter range considered. 
For example, assuming an average transport cost of up to 5  US dollars per barrel of crude oil, the Keystone pipeline in North America operating at full regime (860000 barrels/day) can save about 0.5 million US dollars per day (data from \url{https://en.wikipedia.org/wiki/Keystone_Pipeline, https://en.wikipedia.org/wiki/Petroleum_transport}).

The control becomes effective when the flow spends a significant fraction of the pulsation period in a transient quasi-laminar state, during which the pumping is off, but the flow rate decays slowly and remains high. The time spent by the flow in the quasi-laminar state increases with the forcing period: this is the main reason why long pulsations grant better performance.
This peculiar flow condition, never observed so far, has been proven robust with respect to the discretisation employed. 
Entering the quasi-laminar state clearly requires a low value of the instantaneous $Re$, but this does not limit our approach to low values of the cycle-averaged $Re$. 
The instantaneous value of $Re_m$, in facts, undergoes significant excursions over each forcing period (up to a maximum of $\sim 19000$ in our numerical experiments) and is low only at some time during the cycle. 
Overall, the acceleration intensity and the value of $Re_m$ at the end of the deceleration (which are determined by the choice of the control parameters, $T$ and $\xi$) determine the effectiveness of the control, more than the value of $Re_{m,0}$ itself.

The obtained energy and time savings are robust and significant enough to prove the fundamental result that the “on–off” pumping can be energetically advantageous. 
A practical implementation of the concept will necessarily bring in a process with less-than-unitary efficiency, and consequently extra losses.
They are neglected by this study, which only provides an ideal upper bound for the savings. However, such extra losses are expected to be significantly smaller than the predicted savings, due to the simplicity of the control itself.
Moreover, the full potential of the approach remains to be assessed: there is a huge parameter space to be investigated, and performance could end up being much better.
We also stress that the “on–off” pumping can be applied in several contexts, pipelines being only one of those. Such a pumping strategy is reasonably easy to implement, and does not require complex additional components.

\section*{Methods}

We numerically solve via direct numerical simulation (DNS) the discretised incompressible mass and momentum balance equations with an efficient in-house code (\textit{CFF-Fujin}, \url{https://groups.oist.jp/cffu/code}, \cite{rosti-etal-2020}), capable of second-order accuracy thanks to central finite-differences in space and Adams-Bashfort’s method in time.
Time integration is performed according to a fractional-step method, under a bounding constraint on the maximum value of the CFL number. 
The Poisson pressure equation is dealt with by means of an efficient spectral solver. 
Mass, momentum and energy are therefore conserved exactly. 
The discretisation points are uniformly distributed along the homogeneous directions, while an hyperbolic tangent stretching function is employed in the wall-normal one.

Several resolutions and sizes of the computational domain have been tested to ensure that our conclusions are independent from discretisation effects.
The fully--resolved DNS discussed here have been carried out on a computational grid characterised by a streamwise spacing of $\Delta x^+ = 6.6$ and spanwise spacing $\Delta z^+ = 3.3$, while the wall-normal spacing varies between $\Delta y^+ = 0.5$ at the wall and $\Delta y^+ =  3.2$ at the centreline, in a domain of size $6 \pi h \times 2h \times 3 \pi h$. 
The resolution adopted allows to capture all the scales of turbulent motion induced by our forcing, at the cost of discretising the balance equations over about 42 millions points.

Viscous, or "plus" units are defined for the reference turbulent channel flow under steady forcing at $Re_{m,0} = 5626$ (or $Re_{\tau 0} = 180$). 
This value corresponds to a steady forcing where the pressure gradient has the same magnitude of the time-averaged pressure gradient of the controlled cases. 
In terms of a conventional channel flow at $Re_{m,0} = 5626$, the spatial resolution employed is rather high. 
It should be pointed out that a clearly defined reference flow is lacking, and that the choice of $Re_{m,0}$ is subjective to some extent; the instantaneous value of $Re_m$ undergoes significant excursions.
Hence, the ultimate check for the adequacy of the grid must involve testing different resolutions. 
Such tests, briefly described in the Supplementary Material, do confirm that the flow physics is grid-independent.

Computations were executed by running the program in parallel on $8192$ cores in the supercomputer Deigo at OIST and in Fugaku at RIST. The present results have also been validated against a well tested spectral code \cite{luchini-quadrio-2006}.

\subsection*{Acknowledgments}

G.F.R., A.M. and M.E.R. acknowledge the support of the Okinawa Institute of Science and Technology Graduate University (OIST) with subsidy funding from the Cabinet Office, Government of Japan. 
The authors acknowledge the computer time provided by the Scientific Computing section of the Research Support Division at OIST and M.E.R. also acknowledges the computational time provided by HPCI on the Fugaku cluster under the grant hp210229.

\subsection{Author contributions}
M.Q. conceived the original idea and A.M. performed the preliminary simulations. M.Q. and M.E.R. planned the research and G.F.R. performed the numerical simulations. All authors analyzed data, outlined the manuscript content and wrote the manuscript.

\subsection{Competing interests}
The authors declare that they have no competing interests.

\subsection{Data availability}
All data needed to evaluate the conclusions are present in the paper and/or the Supplementary Material.
Additional data related to this work are available upon reasonable request to the corresponding authors.

\subsection{Code availability}
The codes used for the present research are standard direct numerical simulation solvers for the Navier--Stokes equations. 
Full details of the codes used for the numerical simulations are provided in the Methods section and references therein.

\bibliographystyle{plainnat}
\bibliography{bibliography.bib}

\pagebreak

\section*{Supplementary material}

To assess the robustness of the results, multiple checks have been carried out throughout our demanding numerical study.
Higher grid resolutions, up to a maximum of $\Delta x ^+ = 2.2, \enspace \Delta z^+ = 1.1, \enspace \Delta y^+ = 0.09 - 1.2$, have been tested to verify that the flow physics observed is independent from the discretisation adopted.
The simulation yielding the most promising outcome has also been repeated on a coarser mesh ($\Delta x ^+ = 13.2, \enspace \Delta z^+ = 6.6, \enspace \Delta y^+ = 0.6 - 4.1$, roughly half of the grid points in each direction). The number of forcing periods adopted in the main study was not varied, to keep the time-averaging error at a constant level.
The variation in the computed energy savings has been found to be lower than 1\%, thus verifying the convergence and the grid-independence of the results.
Furthermore, the computational box has been proven large enough to contain the largest flow structures.

For a quantitative assessment of the performance, the proper figures of merit should be carefully designed, as a clearly defined reference flow to compare with is lacking. In this work, we follow the approach proposed by \citet{frohnapfel-hasegawa-quadrio-2012} to set up a rational comparison.
The averaged energy $E$ per unit area spent to transfer a unit mass of fluid through an infinitesimal section of the channel is 
\[
E=\lim_{n \to \infty}\frac{-2 h \int_{n T} \Pi(t) U_b(t)dt}{2 h  \rho \int_{n T} U_b(t) dt}
\]
where $n$ indicates the $n$-th pumping cycle, $h$ is the channel half-width, $T$ the pulsation period, $\Pi(t)$ the instantaneous streamwise pressure gradient, $U_b(t)$ the instantaneous bulk velocity and $\rho$ the fluid density.
The averaged bulk velocity, similarly, is 
\[
U_b = \lim_{n \to \infty}\frac{1}{n T}\int_{n T} U_b(t) dt .
\]

By adopting a turbulent friction law (e.g. the Blasius correlation), the energy ${E_u}$ needed to attain the same bulk velocity $U_{b,c}$ of the controlled flow with an uncontrolled turbulent flow under steady forcing can be computed.
The subscripts $u$ and $c$ denote, respectively, the uncontrolled and controlled flow cases.
Alternatively, the bulk velocity ${U_{b,u}}$ of an uncontrolled turbulent flow under steady forcing leading to the energetic consumption $E_c$ of the controlled flow can be computed.
Interpreting $h/U_b$ as a measure of time, we can therefore define the energy and time savings, respectively, as $S=(E_u-E_c) / (E_u)$ and $(h/U_{b,u} - h/U_{b,c}) / (h/U_{b,u})$: these are the quantities reported in the main text of this paper, and fully account for the energy spent to accelerate the flow during the pulsed operation of the pump.

\begin{figure}
\centering
\includegraphics[scale=0.35]{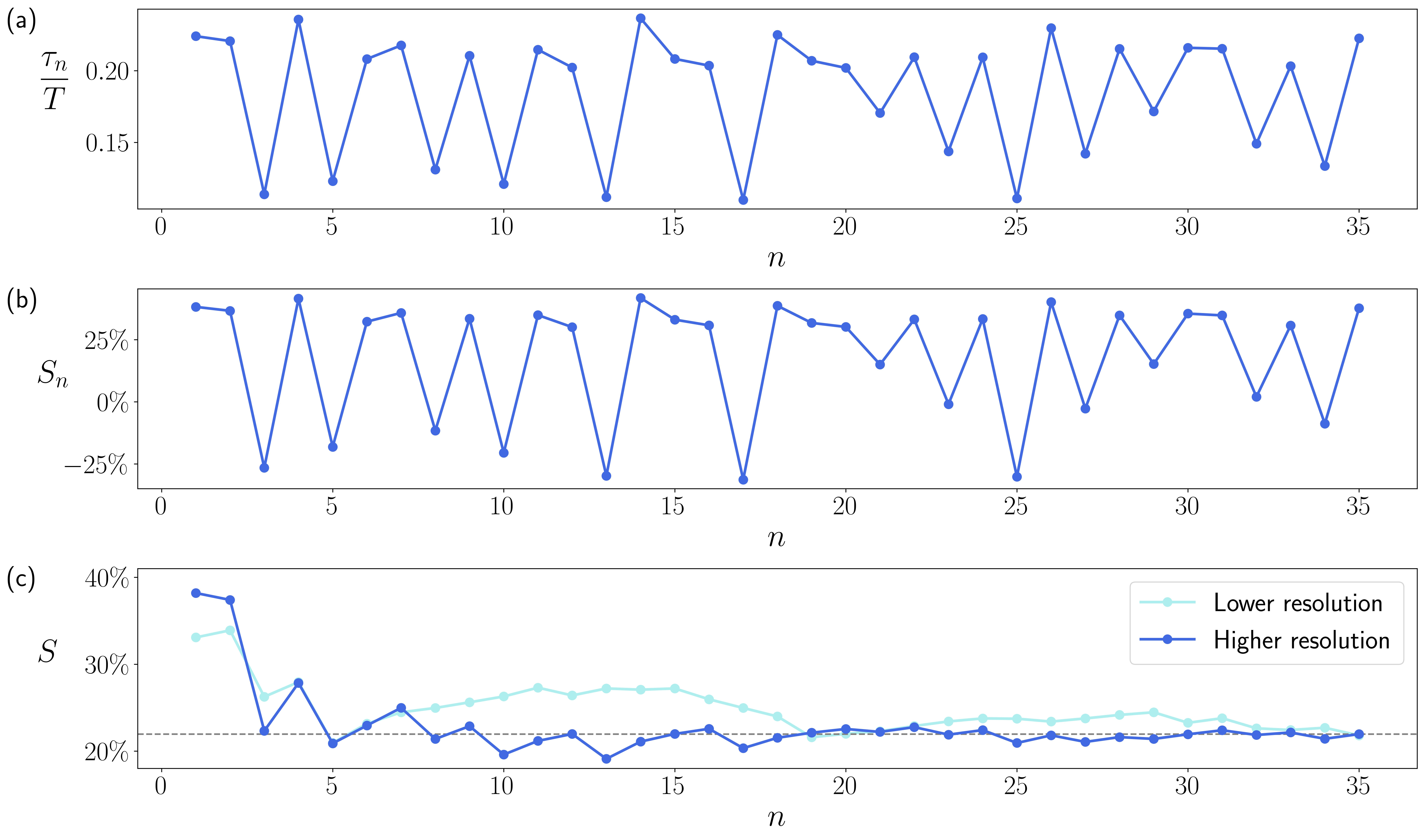}
\caption{\textbf{History and convergence of the energy savings}: we plot the duration of the quasi-laminar flow phase $\frac{\tau_n}{T}$ ($a$) and the energy savings $S_n$ over each forcing period ($b$), accompanied by the the value of a running average of the energy savings $S$ achieved at the end of each period from the beginning of the simulation ($c$). The blue lines refer to the highest resolution considered in the study, the turquoise line to the lower one. $\tau_n$ correlates well with $S_n$, confirming that a longer quasi-laminar flow phase yields higher energy savings. Furthermore, the minimal fluctuations of $S$ over the last periods of the forcing (less than 3\% in relative terms) ensure that its value has reached statistical convergence.}
\label{fig:suppFig}
\end{figure}

When computing $E$ and $U_b$ with a DNS-based numerical experiment, the number of periods available to compute temporal averages is finite. They key random variable affecting convergence of the estimate of the mean saving is the duration of the quasi-laminar phase after pumping is stopped.
In the $n^{th}$ pumping period, the quasi-laminar flow phase lasts for a time $\tau_n$, measured from the end of the acceleration to the appearance of the kink in the time history of the flow rate.
The collection of $\tau_n$ measured in one simulation is a single realisation of the discrete-time stochastic process $\tau$. 
Over a total of 35 periods, we compute a mean value $\overline{\tau}/{T}=0.18$ and a coefficient of variation $CV_{\tau}=0.23$.
$\tau_n$ correlates well with the cycle-averaged energy saving $S_n$ achieved by our control over each forcing period, thus corroborating our physical understanding of the flow dynamics and of the origin of the savings. 
In figure \ref{fig:suppFig} we plot the time histories of $\tau_n$ (panel $a$) and of $S_n$ (panel $b$) for all the periods of the simulation run at the highest resolution.
Panel $c$ also includes data from a lower-resolution simulation, to demonstrate grid independence.
Given the large variance of $\tau$, a robust measurement of the overall energy savings $S$ requires simulations extended over a large number of periods, with the computational burden growing accordingly. 
After 35 periods, results have reached statistical convergence.
As visible in panel $c$ of figure \ref{fig:suppFig}, the running value of $S$ computed at the end of each period is nearly stablized, with a relative fluctuation among the last two periods lower than 3\%.

\end{document}